\documentclass[a4paper,11pt, onecolumn]{article}
\usepackage{amsmath} 		%Advanced Math extensions for LaTex
\usepackage{amssymb} 		%New symbols do math mode
\usepackage{amsthm} 		%Environment proof and theorem
\usepackage{bm} 			%Bold Greek Symbols with command \bm{}
\usepackage{eucal} 			%Other Mathematical symbols
\usepackage{latexsym} 		%Other Mathematical symbols
\usepackage{mathrsfs} 		%Other Mathematica symbols
\usepackage{stmaryrd} 		%More Mathematical symbols
\usepackage{textcomp} 		%Extra Symbols
\usepackage{theorem} 		%Change the style of newly theorems
\usepackage{siunitx} 		%To give the correct spacing in the SI units
\usepackage{slashed} 		%For Feynman slashed notation
\usepackage[official]{eurosym} 	%For the official sign of the euro

\usepackage{array} 			%Extends the possibility to use tables
\usepackage{booktabs} 		%Publication quality tables
\usepackage{caption} 		%provide many ways to customise the captions in floating environments such figure and table
\usepackage{float} 			%Improves the interface for defining floating objects such as figures and tables
\usepackage{graphics}		%to manage external pictures
\usepackage{graphicx} 		%improve graphics
\usepackage{longtable} 		%Multi-pages tables
\usepackage{lscape} 		%requires graphics
\usepackage{subfigure} 		%inclusion of small, `sub,' figures and tables.
\usepackage{wrapfig} 		%To insert pictures wraped in text

\usepackage{ae} 			%Emutates T1 coded fonts
\usepackage[english]{babel} 	%Obligatory. Give the main language to use: english, portuguese
\usepackage[active]{srcltx} 	%DVI search

\usepackage{color} 			%Support for colored text
\usepackage[a4paper]{geometry}		%To set up margins
\usepackage[parfill]{parskip}    % Activate to begin paragraphs with an empty line rather than an indent
\usepackage{rotating} 		%To rotate any kind of objects
\usepackage{setspace} 		%To set the space between lines with commands \onehalfspacing or \doublespacing

\usepackage{cite} 			%Assists in citation management
\usepackage{axodraw4j}
\usepackage{pstricks}

\usepackage{acronym} 		%This package makes sure, that all acronyms used in the text are spelled out in full at least once. And it provides an environment to keep a list of acronyms. Somewhere in the document. 
\usepackage[toc,page,title,header]{appendix} 	%Extra appendices facilities
\usepackage{hyperref} 		%to manage links in the pdf file
\usepackage{makeidx} 		%Activar para fazer um índice remissivo
\usepackage{epstopdf} 		%Conversion
\usepackage{feynmf}

\title{\bf Medium-induced emissions of hard gluons}
\date{\normalsize \today}

\author{{\normalsize Liliana Apolin\'ario$^{a,b}$, N\'estor Armesto$^a$ and Carlos A. Salgado$^a$} \\ \\ \\
{\normalsize \it $^a$ Departamento de F\'{\i}sica de Part\'{\i}culas and  IGFAE,}\\
{\normalsize \it Universidade de Santiago de Compostela, E-15782 Santiago de Compostela, Galicia-Spain}\\ \\
{\normalsize \it $^b$ CENTRA, Instituto Superior T\'ecnico, Universidade T\'ecnica de Lisboa,} \\
{\normalsize \it Av. Rovisco Pais, P-1049-001
Lisboa, Portugal}\\ \\ \\
{\normalsize E-mails: {\tt lilianamarisa.cunha@usc.es, nestor.armesto@usc.es,}}\\ {\normalsize {\tt carlos.salgado@usc.es}.}
}

\begin{document}
\setlength{\parindent}{12pt}

	\maketitle

\begin{abstract}
We present a  derivation of the medium-induced gluon radiation spectrum beyond the current limitation of soft gluon emission. Making use of the path integral approach to describe the propagation of high-energy particles inside a medium, we study the limiting case of a hard gluon emission. %This result is extended to include also the soft gluon emission limit.
Analytical and numerical results are presented and discussed within the multiple soft scattering approximation. An ansatz interpolating between soft and hard gluon emissions is provided. The Landau-Pomeranchuk-Migdal effect is observed in the expected kinematic region.
\end{abstract}

\section{Introduction}

\par In ultrarelativistic heavy-ion collisions, a hot and dense medium is produced. One of the most important probes of this medium is the production of hadrons and jets with high transverse momentum that  undergo energy loss processes. This leads to a collection of phenomena, usually referred to as jet quenching (reviews of the main models can be found in \cite{JQEnterria, JQMajumder, Brick}), which have been already observed in heavy-ion collisions at the Relativistic Heavy Ion Collider (RHIC) ($\sqrt{s_{NN}} = 200$ GeV), such as the suppression of high-$p_T$ hadron spectra, and of di-hadron and $\gamma$-hadron correlations, see e.g. \cite{JQEnterria, JQMajumder,Phenix_pt,Phenix_dihadron,Star_gamma} and refs. therein.

\par Now, with the  heavy-ion program at the Large Hadron Collider (LHC), the current Pb-Pb collision energy is much larger ($\sqrt{s_{NN}} = 2.76$ TeV) and, thus, observables characterized by semi-hard and hard scales are produced more abundantly. Some striking results have already been measured: the observation of a strong asymmetry in dijet transverse energy \cite{ATLAS, CMS,:2012ni},  and a strong nuclear suppression of high-$p_T$  light \cite{ALICE,:2012nt} and heavy \cite{Collaboration:2012nj} hadrons. The magnitude of these features at high energies strongly indicates  the importance of a good description of high transverse momentum observables like jets.  

\par The most commonly used jet quenching models describe the changes in hadron spectra and jet properties in terms of radiative energy loss processes
%\cite{JQEnterria, JQMajumder, Brick}).
\cite{Baier:1996sk,Zakharov:1997uu,Wiedemann:2000za,Gyulassy:2000er,Wang:2001if, Arnold:2002ja}.
This kind of approaches neglects the energy transferred to the medium (recoil) and usually works in the limit of soft gluon emissions i.e.  calculations of the medium-induced gluon  radiation spectrum are  made in the limiting case where the gluon carries a  small fraction $x$ of the energy of its parent parton ($x \rightarrow 0$). The implications of such limitation on the computation of physical observables are discussed at length in \cite{Brick}. Nevertheless, the original techniques in \cite{Baier:1996sk,Zakharov:1997uu,Wiedemann:2000za} can be extended beyond the limit $x\to 0$. This extension was studied for the case of the energy distribution in the BDMPS limit\footnote{We define  the BDMPS limit as the one where $L\to \infty$ with $\omega/\omega_c$ finite, and $\omega_c=\frac{1}{2}\hat q L^2$, see \cite{ASW1} for details. Here  $\hat q$ is the transport coefficient encoding the properties of the medium, $L$ its length and $\omega$ the energy of the radiated gluon. Formally, this limit is equivalent to integrating the gluon transverse momentum up to infinity, neglecting the kinematic constrains.} in Refs. \cite{Zakharov:1997uu,Zakharov,Baier:1998kq,Arnold:2002ja,Arnold:2009mr}. Here we will present results beyond the BDMPS limit for the case of a hard gluon emission;
some attempts in a similar direction can be found in \cite{D'Eramo:2010ak,Ovanesyan:2011xy}.

\par In this work, we use the path-integral formalism  \cite{Zakharov,Baier:1998kq,Wiedemann:2000za,ASW1, ASW2,Kovner:2003zj,lectures} to compute a finite-energy correction  for hard gluon emissions off a quark inside a medium by studying the limiting case  $x \rightarrow 1$. Afterwards, we generalize the result, in a heuristic manner, to include the already known limit of soft gluon emission. The paper is organized as follows. In the next section we will provide a small review of the medium-induced gluon radiation spectrum in the limit of soft gluon emissions, as well as the formalism used. Our calculation of the finite-energy correction will also be made in this section. In section \ref{MSS} we will present the analytic results within the multiple soft scattering approximation \cite{Zakharov,Baier:1996sk} for the final interpolation expression and the numerical discussion will be made in section \ref{sec:results}. Finally, in section \ref{sec:conclusions} a summary and future prospects will be presented. 

Let us note that information about the size of finite energy corrections to medium-induced gluon radiation is badly needed in order to put the existing Monte Carlo models for jet quenching, which automatically include energy-momentum conservation - see e.g. \cite{qpythia1, qpythia2,Zapp:2012nw}, on more solid grounds. Together with research on the role of interferences from different emitters in the QCD shower \cite{MehtarTani:2010ma,MehtarTani:2011tz,CasalderreySolana:2011rz,Armesto:2011ir,MehtarTani:2011gf}, and the effects of color exchanges with the medium \cite{Beraudo:2011bh,Aurenche:2011rd}, they constitute one of the main avenues in the recent developments of the theory of energy loss processes in a QCD medium.

\section{Medium-induced gluon radiation	}

%***********************************
% Image 
%***********************************
\begin{figure}[htbp]
	\begin{center}
		\fcolorbox{white}{white}{
%			\begin{picture}(102,43) (15,-30)
                            \begin{picture}(142,43) (55,-30)
				\scalebox{0.5}{
					\SetWidth{1.0}
					\SetColor{Black}
					\GOval(80,-32)(32,32)(0){0.882}
					\SetWidth{2.0}
					\Line[arrow,arrowpos=0.5,arrowlength=11.562,arrowwidth=4.625,arrowinset=0.2](112,-32)(224,-32)
					\Text(144,-16)[lb]{\Huge{\Black{$p$}}}
					\Line[arrow,arrowpos=0.5,arrowlength=11.562,arrowwidth=4.625,arrowinset=0.2](224,-32)(304,-80)
					\Gluon(224,-32)(304,16){7.5}{5}
					\Text(314,16)[lb]{\Huge{\Black{$ k= x p$}}}
					\Text(320,-96)[lb]{\Huge{\Black{$ q = (1-x) p$}}}
					\Text(64,-37)[lb]{\Huge{\Black{$M_h$}}}
    				}
			\end{picture}
		}
		\vskip 0.5cm
		\caption{Radiation process of a gluon off a quark with the corresponding kinematical variables.}
		\label{fig:Inel}
	\end{center}
\end{figure}
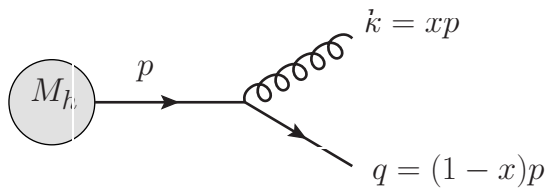

\par Consider the inelastic process shown in figure \ref{fig:Inel}. A quark of 4-momentum\footnote{We will use light-cone coordinates defined as $a=(a_0,a_x,a_y,a_z)=(a_+,a_-,{\bf a}_\perp)$ with $a_\pm=(a_0\pm a_z)/\sqrt{2}$ and ${\bf a}_\perp =(a_x,a_y)$. Furthermore, $C_A=N$ and $C_F=(N^2-1)/(2N)$ are the quadratic Casimirs in the adjoint and fundamental representations respectively, with $N$ the number of colors.}
$p$, ${\bf p}_\perp=0$, coming from a hard process with amplitude $M_h$, emits a gluon of 4-momentum $k$ with $+$-component  $k_+=x p_+$ and transverse momentum ${\bf k}_\perp$.
%
%$p = (p, 0, \mathbf{0_\perp} )$ coming from a hard process with amplitude $M_h$ emits a gluon of momentum $k = \left( \omega , 0, \mathbf{k_\perp} \right) = \left( \sqrt{2} x p , 0, \mathbf{k_\perp} \right)$ carrying a fraction $x$ of the total initial energy.
%
\par Inside a medium, partons undergo multiple scattering. In the high-energy limit, the parton essentially conserves its energy and the effect of the medium is only a rotation of its color field, thus acquiring an eikonal phase. In this situation, a  convenient way of describing the parton propagation is using Wilson lines,
%-----------------------------------
% Equation
%-----------------------------------
\begin{equation}
	W ( x_{0+}, L_+; \mathbf{x_\perp} ) = \mathcal{P}  \exp \left\{ i g \int_{x_{0+}}^{L_+} d x_+ A_- (x_+, \mathbf{x_\perp} ) \right\},
	\label{eq:WL}
\end{equation} 
which describe the propagation of a particle through a medium with longitudinal boundaries at $ \left[ x_{0+}, L_+ \right] $ and color field $A_-$. As for the less energetic parton, the restrictions of the above formula need to be relaxed to allow some motion in the transverse plane of the propagating particle (see e.g. \cite{Kovner:2003zj,lectures} and refs. therein). By doing this, the Wilson line is replaced by the path integral propagator from $x_+, {\bf x}_{0\perp}$ to $L_+, {\bf x}_{\perp}$,
%-----------------------------------
% Equation
%-----------------------------------
\begin{equation}
\begin{split}
	G ( x_{0+}, \mathbf{x_{0\perp}}; L_+, \mathbf{x_\perp} | p_+) & = \int_{ \mathbf{r_\perp} (x_{0+}) = \mathbf{x_{0\perp}} }^{ \mathbf{r_\perp} (L_+) = \mathbf{x_\perp} } \mathcal{D} \mathbf{r_\perp} (\xi) \exp \left\{ \frac{ i p_+}{ 2} \int_{x_{0+}}^{L_+}  d\xi \; \left( \frac{ d\mathbf{r_\perp}}{ d\xi} \right)^2 \right\}  \\
	& \times W( x_{0+}, L_+; \mathbf{r_\perp} (\xi) ) ,
\end{split}
	\label{eq:Green}
\end{equation}
which associates a Brownian motion of the particle in the transverse plane at the same time that its color field is rotated. %The boundary conditions on the free path integral are $\mathbf{r_\perp} (x_{0+}) = \mathbf{x_{0\perp}}$ and $\mathbf{r_\perp} (L_+) = \mathbf{x_\perp}$. 

\subsection{Soft limit: $\mathbf{ x \rightarrow 0} $ }

\par
\begin{figure}[htbp]
	\begin{center}
		\subfigure[Radiation vertex outside the medium.]{
			\fcolorbox{white}{white}{
				\begin{picture}(151,57) (15,-25)
					\scalebox{0.35}{
						\SetWidth{1.0}
						\SetColor{Black}
						\GOval(80,10)(32,32)(0){0.882}
						\SetWidth{2.0}
						\Line[arrow,arrowpos=0.865,arrowlength=11.562,arrowwidth=5,arrowinset=0.2](112,10)(448,10)
						\Text(208,26)[lb]{\Huge{\Black{$p$}}}
						\Gluon(336,10)(432,74){7.5}{6}
						\Text(448,74)[lb]{\Huge{\Black{$k$}}}
						\Text(464,10)[lb]{\Huge{\Black{$q \simeq p$}}}
						\SetWidth{2.0}
						\Gluon(160,10)(160,-70){7.5}{4}
						\Vertex(160,-70){4}
						\Vertex(272,-70){4}
						\Gluon(272,10)(272,-70){7.5}{4}
						\Text(210,-38)[lb]{\Huge{\Black{$...$}}}
					}
				\end{picture}
			}
			\label{fig:x0q}
		}
		\hfill
		\subfigure[Radiation vertex inside the medium.]{
			\fcolorbox{white}{white}{
				\begin{picture}(193,73) (15,-9)
					\scalebox{0.35}{
						\SetWidth{1.0}
						\SetColor{Black}
						\GOval(80,58)(32,32)(0){0.882}
						\SetWidth{2.0}
						\Line[arrow,arrowpos=0.925,arrowlength=11.562,arrowwidth=5,arrowinset=0.2](112,58)(576,58)
						\Text(210,74)[lb]{\Huge{\Black{$p$}}}
						\Gluon(336,58)(560,154){7.5}{13}
						\Text(576,154)[lb]{\Huge{\Black{$k$}}}
						\Text(592,58)[lb]{\Huge{\Black{$q \simeq p$}}}
						\SetWidth{2.0}
						\Gluon(160,58)(160,-22){7.5}{4}
						\Vertex(160,-22){4}
						\Vertex(272,-22){4}
						\Gluon(272,58)(272,-22){7.5}{4}
						\Text(210,10)[lb]{\Huge{\Black{$...$}}}
						\Gluon(496,58)(496,-22){7.5}{4}
						\Gluon(400,58)(400,-22){7.5}{4}
						\Vertex(496,-22){4}
						\Vertex(400,-22){4}
						\Text(400,132)[lb]{\Huge{\Black{$...$}}}
						\Text(442,10)[lb]{\Huge{\Black{$...$}}}
						\Vertex(448,186){4}
						\Vertex(352,154){4}
						\Gluon(384,90)(352,154){7.5}{4}
						\Gluon(480,122)(448,186){7.5}{4}
					}
				\end{picture}
			}
			\label{fig:x0g}
		}
		\label{fig:x0}
		\caption{Radiation diagrams in the limiting case $x \rightarrow 0$.}
	\end{center}
\end{figure}
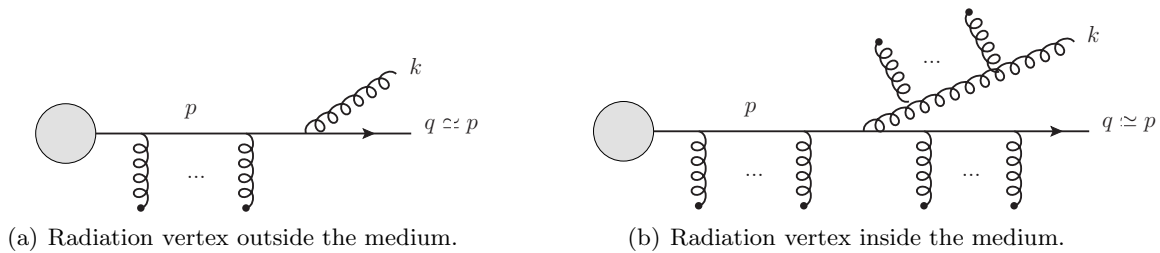

 Considering the soft gluon emission limit ($x \rightarrow 0 $), there are two contributions for the total amplitude: the case where the gluon does not interact after it is emitted (figure \ref{fig:x0q}), and the case where the radiation vertex is inside the medium (figure \ref{fig:x0g}).

%***********************************
% Image 
%***********************************

This means that the total amplitude is given by
%-----------------------------------
% Equation
%-----------------------------------
\begin{equation}
	M_{tot}^{x \rightarrow 0} = M_q^{x \rightarrow 0} + M_g^{x \rightarrow 0},
\end{equation}
where $ M_q^{x \rightarrow 0}$ describes the diagram of figure \ref{fig:x0q} and $ M_g^{x \rightarrow 0}$ describes the diagram of figure \ref{fig:x0g}. Thus, the double differential spectrum of medium-induced radiation in this limit reads
%-----------------------------------
% Equation
%-----------------------------------
\begin{equation}
	\left. k_+ \frac{ dI^{tot} }{ d k_+ d^2 \mathbf{k_\perp} } \right|_{x \rightarrow 0} = \frac{Ê\left\langle \overline{ | M_{tot}^{x \rightarrow 0 } |^2 } \right\rangle }{ 2 (2 \pi)^3 } =  \left. k_+ \frac{ dI^{vac} }{ d k_+ d^2 \mathbf{k_\perp} }\right|_{x \rightarrow 0} +  \left.k_+ \frac{ dI^{med} }{ d k_+ d^2 \mathbf{k_\perp} }\right|_{x \rightarrow 0}\,,
\end{equation}
where $\left\langle \cal O \right\rangle$ denotes the medium average of ${\cal O}$, see e.g.  \cite{Kovner:2003zj,lectures} and $\overline{\cal O}$ makes reference to the spin and color averages. The vacuum contribution is recovered from $\overline{ \left\langle | M_{q}^{x \rightarrow 0 } |^2 \right\rangle} $:
%----------------------------------- 
% Equation
%-----------------------------------
\begin{equation}
	\left. k_+ \frac{ dI^{vac} }{ d k_+ d^2 \mathbf{k_\perp} } \right|_{x \rightarrow 0} = \frac{ \alpha_s C_FÊ}{ \pi^2 } \frac{Ê1}{ \mathbf{k_\perp^2 } }\,.
\end{equation}

The medium contribution comes from the diagram \ref{fig:x0g} with its own complex conjugate ($\left\langle \overline{ | M_{g}^{x \rightarrow 0 } |^2} \right\rangle $), usually called \textit{gluon} term, and from this same diagram with the complex conjugate of diagram \ref{fig:x0q} ($ \left\langle \overline{ M_{g}^{x \rightarrow 0 } \left( M_{q}^{x \rightarrow 0 } \right)^\dagger } \right\rangle $ ), denoted  \textit{interference} term\footnote{Full explanations of the variables in this equation can be found in \cite{ASW1, ASW2,Kovner:2003zj,lectures}.}:
%-----------------------------------
% Equation
%-----------------------------------
\begin{equation}
\begin{split}
	\left. k_+ \frac{ dI^{med} }{ d k_+ d^2 \mathbf{k_\perp} } \right|_{x \rightarrow 0} & = \frac{ \alpha_s  C_F }{ (2\pi)^2 k_+ } 2 \text{Re} \left\{ \frac{ 1Ê}{ k_+ } \int dy_+ d\bar{y}_+ d \mathbf{x_\perp} \text{e}^{- i \mathbf{k_\perp} \cdot \mathbf{x_\perpÊ} } \text{e}^{- \frac{1}{2} \int d \xi n(\xi) \sigma (\mathbf{x_\perp}) } \right. \\
	&\times  \left. \frac{ \partial }{ \partial \mathbf{y_\perp} } \cdot \frac{ \partial }{ \partial \mathbf{x_\perpÊ} } \mathcal{K} (y_+, \mathbf{y_\perp} = \mathbf{0}; \bar{y}_+, \mathbf{x_\perp}|k_+ ) \right. \\
	& \left. + \int dy_+ d\mathbf{x_\perp} \text{e}^{- i \mathbf{x_\perp} \cdot \mathbf{k_\perp} } \ 2 \frac{ \mathbf{k_\perp} }{ \mathbf{k_\perp}^2Ê} \cdot \frac{ \partial }{ \partial \mathbf{y_\perp} } \mathcal{K} (y_+, \mathbf{y_\perp} = \mathbf{0}; L_+, \mathbf{x_\perp} |k_+) \right\}.
\end{split}
\label{eq:xto0}
\end{equation}
%{\bf CHECK THE $\omega$'s IN THE LHS}\\
Here, $\mathcal{K}$ denotes the two-dimensional path-integral
%-----------------------------------
% Equation
%-----------------------------------
\begin{eqnarray}
	\mathcal{K} (x_+, \mathbf{x_\perp}; y_+, \mathbf{y_\perp} |k_+)&=& \int_{ \mathbf{r_\perp} (x_+) = \mathbf{x_\perp} }^{\mathbf{r_\perp} (y_+) = \mathbf{y_\perp} } \mathcal{D} \mathbf{r_\perp} (\xi) \\
	&\times& \exp \left\{ \int_{x_+}^{y_+} d\xi \left[ \frac{ i k_+ }{ 2} \left( \frac{ d\mathbf{r_\perp}}{ d\xi} \right)^2
	 - \frac{ 1}{2} n(\xi) \sigma \left( \mathbf{r_\perp } \right) \right] \right\}.\nonumber 
\end{eqnarray}
This results from the fact that one has to average over all medium configurations since for the calculation, only a frozen configuration profile was taken into account. The medium dependence of the spectrum comes from the factor $n (\xi) \sigma (\mathbf{r_\perp} )$. The density of scattering centers is given by $n(\xi)$ and their space configuration and strength is contained in the dipole cross section,
%-----------------------------------
% Equation
%-----------------------------------
\begin{equation}
	\sigma (\mathbf{r_\perp} ) = 2
	%C_F
	\int \frac{ d \mathbf{q_\perp} }{ (2\pi)^2 } \left| a_0 (\mathbf{q_\perp} ) \right|^2 \left( 1 - \text{e}^{- i \mathbf{q_\perp} \cdot \mathbf{r_\perp} } \right),
	\label{eq:DipoleX}
\end{equation}
with $a_0 (\mathbf{q_\perp} )$ the scattering potential corresponding to one scattering center, usually taken in the form of a static Debye screened potential \cite{Gyulassy:1993hr}.

%\textbf{Check the definition of $\sigma$}

\subsection{Hard limit:  $\mathbf{x \rightarrow 1} $  }
\label{sec:x1}

\par We will now to parallel the derivation in the previous section but in the limit $x \rightarrow 1$. This means that the kinematics of the process is now constrained by
%-----------------------------------
% Equation
%-----------------------------------
\begin{equation}
	|\mathbf{q_{\perp}}|\ll q_+ \ll k_+, p_+\,.
\end{equation}

\par We work in the high-energy limit, which means that terms proportional to $\mathbf{p_{i_\perp}}$, $\mathbf{q_{i_\perp}}$ or $\mathbf{k_{i_\perp}}$ are neglected in the numerator of the propagators \cite{Kovner:2003zj,lectures}. As for the denominators, the propagators coming from the initial quark and from the gluon will be simplified to
%-----------------------------------
% Equation
%-----------------------------------
\begin{equation}
	\frac{ i \slashed{p}_i }{ p_i^2 + i \varepsilon} \simeq \frac{ i \slashed{p}_i }{ 2 p_{i_+} p_{i_-} + i \varepsilon}\,,
\end{equation}
whereas for the final quark, the term proportional to $\mathbf{q}_{i\perp}^2$ is kept in the denominator:
%-----------------------------------
% Equation
%-----------------------------------
\begin{equation}
	\frac{ i \slashed{q}_i }{ q_i^2 + i \varepsilon} \simeq \frac{ i \slashed{q}_i }{ 2 q_{i_+} q_{i_-} - {\bf q}_{i\perp}^2 + i \varepsilon}\,.
\end{equation}
By doing this we are assuming that the most energetic partons (the initial quark and the gluon) will  acquire only a color phase by crossing the medium. Since the final quark is much  softer than the other two, the contribution from $\mathbf{q_{i\perp}}$ in the denominator must be taken into account to allow some motion in the transverse plane.

%***********************************
% Image 
%***********************************
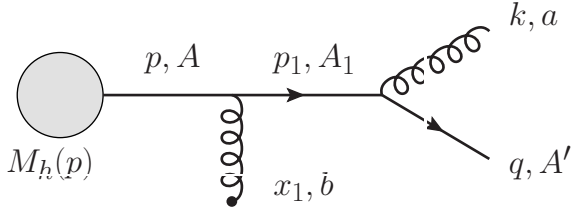
\begin{figure}[tp]
	\begin{center}
		\fcolorbox{white}{white}{
			\begin{picture}(145,51) (25,-30)
				\scalebox{0.5}{
					\SetWidth{1.0}
					\SetColor{Black}
					\GOval(80,-6)(32,32)(0){0.882}
					\SetWidth{2.0}
					\Line[arrow,arrowpos=0.69,arrowlength=11.562,arrowwidth=4.625,arrowinset=0.2](112,-6)(320,-6)
					\Text(144,10)[lb]{\Huge{\Black{$p, A$}}}
					\Text(240,10)[lb]{\Huge{\Black{$p_1, A_1$}}}
					\Line[arrow,arrowpos=0.5,arrowlength=11.562,arrowwidth=4.625,arrowinset=0.2](320,-6)(400,-54)
					\Gluon(320,-6)(400,42){7.5}{5}
					\Text(416,42)[lb]{\Huge{\Black{$k, a$}}}
					\Text(416,-70)[lb]{\Huge{\Black{$q, A'$}}}
					\SetWidth{2.0}
					\Gluon(208,-6)(208,-86){7.5}{4}
					\Vertex(208,-86){4}
					\Text(240,-86)[lb]{\Huge{\Black{$x_1, b$}}}
					\Text(40,-70)[lb]{\Huge{\Black{$M_h (p)$}}}
				}
			\end{picture}
		}
		\vskip 0.5cm
		\caption{Radiation diagram for gluon emission outside the medium in the one-scattering case. $p,p_1,k,q$ denote the 4-momenta, $x_1$ the position of the scattering center, and $A,A_1,A^\prime$ the color indices in the fundamental representation, while $a,b$ denotes the color index in the adjoint.}
		\label{fig:x1Mq1}
	\end{center}
\end{figure}

\par First, we will compute the contribution where the gluon is emitted outside the medium. Starting by one scattering with the medium (diagram represented in figure \ref{fig:x1Mq1} ), the ${\cal T}$-matrix reads
%-----------------------------------
% Equation
%-----------------------------------
\begin{equation}
	{\cal T}_1^q = \int \frac{Êd^4p}{ (2\pi)^4Ê} d^4x_1 \frac{ d^4 p_1}{(2\pi)^4} \text{e}^{i x_1 \cdot (p_1-p) } \bar{u} (q) ig T_{A'A_1}^{a} \slashed{\epsilon}^* \frac{ i \slashed{p}_1}{ p_1^2 + i \varepsilon } ig \slashed{A}_{A_1 A} (x_{1_+}, \mathbf{x_{1\perp}} ) \frac{ i \slashed{p} }{ p^2 + i \varepsilon} M_h (p),
	\label{eq:Sq1}
\end{equation}
with $A\equiv A^b T^b$.
For the simplification of the Dirac structure, the approximations used in the soft gluon emission cannot be applied \footnote{In the case $x \rightarrow 0$, $q\simeq p$ and thus the following simplification was used: $\bar{u} (q) \slashed{\epsilon}^* \slashed{p}_1 \simeq \bar{u} (q) \slashed{\epsilon}^* \slashed{q} \simeq 2 q \cdot \epsilon^* \simeq 2 q_+ \epsilon^*_-$, where $\epsilon =  \left( \epsilon_+, \epsilon_-, \boldsymbol{\epsilon_\perp} \right) $ is the gluon polarization. 
Now, for $x \rightarrow 1$, the momenta relation has changed and the Dirac equation can no longer be used. Besides, we have to keep $\slashed{\epsilon}^*$ unevaluated since we do not know a priori which component will be dominant as $k_+$ is larger than in the previous case. This means that the Dirac structure can only be reduced to $\bar{u} (q) \slashed{\epsilon}^* (\slashed{q} + \slashed{k} ) \slashed{A}^b_{A_1 A} \slashed{p} M_h (p) = 2 (q+k)_+ (A_-)_{A_1 A} p_+ \bar{u} (q) \slashed{\epsilon}^* \gamma_- M_h (p)$ (see appendix \ref{appA} for the relations between the $\gamma$ matrices).}. As for the integrals, the only non-trivial ones  are the ones related with the propagators:
%-----------------------------------
% Equation
%-----------------------------------
\begin{equation}
	\int \frac{ dp_-}{ 2\pi} \text{e}^{-i p_- (x_{(i+1)+} - x_{i+})} \frac{ i }{ p_- - (- i \varepsilon)} = \Theta (x_{(i+1)+} - x_{i+} ),
\end{equation}
where $\Theta$ is the step function. Thus equation \eqref{eq:Sq1} can be reduced to

%-----------------------------------
% Equation
%-----------------------------------
\begin{equation}
	{\cal T}_1^q = - g T_{A' A_1}^a \frac{ (q+k)_+ }{ 2 (q \cdot k) } \int dx_{1_+}\left[ i g (A_-) _{A_1 A} (x_{1+}, \mathbf{0_\perp} )\right] \Theta (x_{1+}) \bar{u} (q) \slashed{\epsilon}^* \gamma_- M_h(k+q).
\end{equation}

Generalizing to $n$ scatterings, one can check that this structure iterates. Summing over all scattering centers an exponential series is found and the  ${\cal T}$-matrix for a gluon emitted outside the medium can be written as follows:
%-----------------------------------
% Equation
%-----------------------------------
\begin{equation}
	{\cal T}_q = - g T_{A' A_1}^a \frac{ (q+k)_+ }{ 2 (q \cdot k) } W_{A_1 A} (x_{0+}, L_+; \mathbf{0}_\perp ) \bar{u} (q) \slashed{\epsilon}^* \gamma_- M_h(k+q).
	\label{eq:Sq}
\end{equation}

\begin{figure}[tp]
	\begin{center}
		\fcolorbox{white}{white}{
			 \begin{picture}(220,99) (16,-4)
				\scalebox{0.4}{
					\SetWidth{1.0}
					\SetColor{Black}
					\GOval(80,138)(32,32)(0){0.882}
					\SetWidth{2.0}
					\Line[arrow,arrowpos=0.5,arrowlength=11.562,arrowwidth=4.625,arrowinset=0.2](112,138)(384,138)
					\Text(128,154)[lb]{\Huge{\Black{$p, A$}}}
					\Line[arrow,arrowpos=0.575,arrowlength=11.562,arrowwidth=4.625,arrowinset=0.2](384,138)(656,42)
					\Gluon(384,138)(656,234){7.5}{15}
					\Text(672,234)[lb]{\Huge{\Black{$k, a$}}}
					\Text(672,42)[lb]{\Huge{\Black{$q, A'$}}}
					\SetWidth{2.0}
					\Gluon(192,138)(192,58){7.5}{4}
					\Vertex(496,26){4}
					\Gluon(512,90)(496,26){7.5}{3}
					\Gluon(496,250)(512,186){7.5}{4}
					\Vertex(288,58){4}
					\Vertex(592,282){4}
					\Vertex(192,58){4}
					\Gluon(608,58)(592,-6){7.5}{3}
					\Gluon(592,282)(608,218){7.5}{4}
					\Gluon(288,138)(288,58){7.5}{4}
					\Vertex(592,-6){4}
					\Vertex(496,250){4}
					\Text(230,100)[lb]{\Huge{\Black{$...$}}}
					\Text(64,74)[lb]{\Huge{\Black{$M_h(p)$}}}
					\Text(540,230)[lb]{\Huge{\Black{$...$}}}
					\Text(545,42)[lb]{\Huge{\Black{$...$}}}
					\Text(370,106)[lb]{\Huge{\Black{$y$}}}
					\Text(304,154)[lb]{\Huge{\Black{$p_1, A_1$}}}
					\Text(410,185)[lb]{\Huge{\Black{$k_1, a_1$}}}
					\Text(425,65)[lb]{\Huge{\Black{$q_1, A_1'$}}}
				}
			\end{picture}
		}
		\caption{Radiation diagram for gluon emission inside the medium for the limiting case $x \rightarrow 1$. The meaning of the variables and indices is analogous to that in figure \ref{fig:x1Mq1}.}
		\label{fig:x1Mg}
	\end{center}
\end{figure}
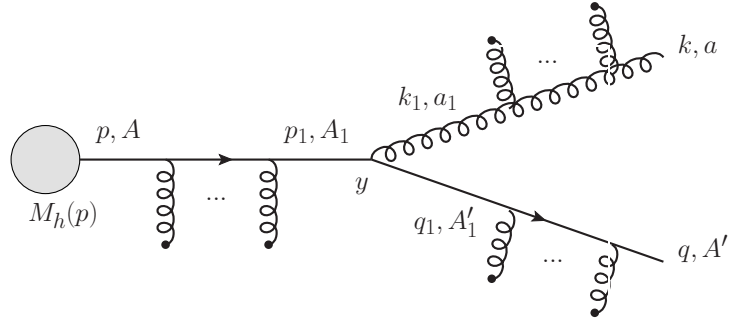

We can repeat the same procedure for the process in which the radiation vertex is inside the medium (see figure \ref{fig:x1Mg}). For one scattering, 
%-----------------------------------
% Equation
%-----------------------------------
\begin{equation}
\begin{split}
	{\cal T}_1^g & = \int \frac{Êd^4p}{ (2\pi)^4Ê} d^4x_1 \frac{ d^4 p_1}{(2\pi)^4} d^4 y \frac{ d^4 q_1 }{ (2\pi)^4} d^4 x_1' \frac{d^4 k_1}{ (2\pi)^4 } d^4y_1 \text{e}^{i x_1 \cdot (p_1-p) } \text{e}^{i x_1' \cdot (q - q_1) } \text{e}^{i y_1 \cdot (k-k_1) } \text{e}^{i y \cdot (q_1+k_1-p_1) } \\
	& \times \bar{u} (q) ig \slashed{A}_{A' A_1'} (x_{1+}', \mathbf{x_{1\perp}'}) \frac{ i \slashed{q}_1}{ q_1^2 + i \varepsilon} ig T_{A_1'A_1}^{a_1} \gamma^{\mu_1} \frac{ i \slashed{p}_1}{ p_1^2 + i \varepsilon } ig \slashed{A}_{A_1 A} (x_{1_+}, \mathbf{x_{1\perp}} ) \frac{ i \slashed{p} }{ p^2 + i \varepsilon} M_h (p) \\
	& \times \epsilon^*_\mu(k) g f^{a a_1 b} V^{\mu \mu_1' \nu} (-k, k_1, 0) A_\nu^b (y_{1+}, \mathbf{y_{1\perp}} ) \frac{ -i g_{\mu_1' \mu_1} }{ k_1^2 + i \varepsilon} \,,
	\label{eq:Sq2}
\end{split}
\end{equation}

\par The gluon vertex \footnote{$V^{\alpha \beta \delta} (k_1, k_2, k_3) = g^{\alpha \beta} (k_1-k_2)^\delta + g^{\beta \delta} (k_2-k_3)^\alpha + g^{\delta \alpha} (k_3 - k_2)^\beta$.}, together with the metric from the propagator and the polarization vector can be simplified using $k_{1+} = k_{+}$, and so, the Dirac structure takes the form:
%-----------------------------------
% Equation
%-----------------------------------
\begin{equation}
	\underbrace{ \bar{u} (q) \slashed{A}_{1}' \slashed{q}_1}_{(b)} \slashed{\epsilon}^* (k) \underbrace{ \slashed{p_1} \slashed{A}_{1} \slashed{p}}_{(a)}
\end{equation}
where
\begin{equation}
	(a) = \gamma_- p_{1+} \slashed{ AÊ}_{1} p_+ \gamma_- = p_{1+} A_{1} p_- \gamma_- \gamma_+ \gamma_- = 2 p_{1+} A_{1} p_+ \gamma_-\,,
\end{equation}
\begin{equation}
	(b) = 2 q_1 \cdot A_{1}' \bar{u} (q) - \underbrace{ \bar{u}Ê(q) \slashed{q}_1 }_{\simeq 0} \slashed{A}_{1}' \,.
\end{equation}
In this last simplification, we are keeping only the dominant term of the Dirac equation since $q_{1+} = q_+$. But we must not forget that the $\mathbf{q_\perp}$ coming from the spinor $\bar{u}(q)$ is actually $\mathbf{q_{1 \perp}} \neq \mathbf{q_{\perp}} $. This means that in the squared modulus, the transverse momentum that appears from this $\mathcal{T}$-matrix corresponds to an inner momentum. The same is applied to the gluon transverse momentum coming from the gluon polarization vector, $\mathbf{k_\perp} = -\mathbf{q_{1\perp}}$.

\par Using the properties listed in appendix \ref{appA} and 
%-----------------------------------
% Equation
%-----------------------------------
\begin{equation}
\begin{split}
	\int \frac{ dq_-}{ 2\pi} & \frac{ d \mathbf{q_\perp} }{ (2\pi)^2 } \text{e}^{-i q_- (x_{(i+1)+} - x_{i+}) + i \mathbf{q_\perp} \cdot ( \mathbf{x}_{(i+1)\perp} - \mathbf{x}_{i\perp} ) } \frac{ i }{ q_- - (q_\perp^2 / 2q_+ - i \varepsilon)}  \\
	& = \Theta (x_{(i+i)+} - x_{i+} ) G_0 (x_{i+}, \mathbf{x}_{i\perp}; x_{(i+1)+}, \mathbf{x}_{(i+1)\perp } | q_+ ),
\end{split}
\label{eq:int_g0}
\end{equation}
%***********************************
% Image 
%***********************************
where
%-----------------------------------
% Equation
%-----------------------------------
\begin{equation}
\begin{split}
	G_0 (x_{i+}, \mathbf{x}_{i\perp}; & x_{(i+1)+}, \mathbf{x}_{(i+1)\perp } | q_+)  = \frac{Êq_+ }{ 2 \pi i (x_{(i+1)} - x_i)_+} \exp \left\{ \frac{ i p_+ }{ 2 } \frac{ ( \mathbf{x}_{(i+1)} - \mathbf{x}_i )_\perp^2 }{ (x_{(i+1)} - x_i)_+ } \right\} \\
	& \equiv \int_{\mathbf{r_\perp} (x_{i+}) = \mathbf{x}_{i\perp}}^{\mathbf{r_\perp} (x_{(i+1)+}) = \mathbf{x}_{(i+1)\perp}} \mathcal{D} \mathbf{r_\perp} (\xi) \exp \left\{ \frac{ i q_+ }{ 2} \int_{x_{i+}}^{x_{(i+1)+}} d\xi \; \left( \frac{ d\mathbf{r_\perp}}{ d\xi } \right)^2 \right\} 
\end{split}
\end{equation}
is the Green's function of a free particle that propagates in the transverse plane from $\mathbf{x}_{i\perp}$ at (light-cone) time $x_{i+}$ to $\mathbf{x}_{(i+1)\perp}$ at time $x_{(i+1)+}$, we get for  the  ${\cal T}$-matrix for a gluon emitted inside the medium
%-----------------------------------
% Equation
%-----------------------------------
\begin{equation}
\begin{split}
	{\cal T}_g & = \frac{ 1Ê}{2 } \int dy_+ d\mathbf{x_\perp} \text{e}^{-i \mathbf{q_\perp} \cdot \mathbf{x_\perp} } G_{A' A_1'} (y_+, \mathbf{y_\perp} = \mathbf{0}_\perp; L_+, \mathbf{x_\perp} | q_+ ) ig T^{a_1}_{A_1' A_1} \\ 
	& \times  W_{A_1 A} (x_{0+}, y_+;  \mathbf{0}_\perp ) W_{a a_1} (y^+, L^+; \mathbf{0}_\perp ) \bar{u} (q) \slashed{\epsilon}^* (k) \gamma_- M_h (q+k),
	\label{eq:Sg}
\end{split}
\end{equation}
where the use of uppercase (lowercase) color indices in the Wilson lines indicate that they are to be taken in the fundamental (adjoint) as they correspond to the rescattering of a quark (gluon).

% and we have also used that
%
%%-----------------------------------
%% Equation
%%-----------------------------------
%\begin{equation}
%\begin{split}
%	\boldsymbol{\epsilon_\perp}^* \cdot \mathbf{k}_{1\perp} G_0 (y_+, \mathbf{y_{\perp}} = \mathbf{0}_\perp; & x_{1+}'; \mathbf{x}_{1\perp} | q_+) = - \boldsymbol{\epsilon_\perp}^* \cdot \mathbf{q}_{1\perp} G_0 (y_+, \mathbf{y_{\perp}} = \mathbf{0}+\perp; x_{1+}'; \mathbf{x}_{1\perp} | q_+) \\
%	& = - i \boldsymbol{\epsilon_\perp}^* \cdot \frac{ \partial }{ \partial \mathbf{y_{\perp}}} G_0 (y_+, \mathbf{y_{\perp}} = \mathbf{0}_\perp; x_{1+}'; \mathbf{x}_{1\perp} | q_+).
%\end{split}
%\end{equation}

\par
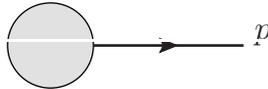
\begin{figure}[bp]
	\begin{center}
		\fcolorbox{white}{white}{
			\begin{picture}(86,22) (30,-42)
				\scalebox{0.5}{ 
					\SetWidth{1.0}
					\SetColor{Black}
					\GOval(80,-94)(32,32)(0){0.882}
					\SetWidth{2.0}
					\Line[arrow,arrowpos=0.5,arrowlength=12.5,arrowwidth=5,arrowinset=0.2](112,-94)(224,-94)
					\Text(233,-94)[lb]{\Huge{\Black{$p$}}}
				}
			\end{picture}
		}
		\vskip 0.8cm
		\caption{Elastic process.}
		\label{fig:elastic}
	\end{center}
\end{figure}

 The total ${\cal T}$-matrix, ${\cal T}_{tot}$, is the sum of both contributions (equations \eqref{eq:Sq} and \eqref{eq:Sg}). The spectrum is computed as the inelastic cross-section over the elastic cross-section (see the elastic process in figure \ref{fig:elastic}).
%***********************************
% Image 
%***********************************
Thus
%-----------------------------------
% Equation
%-----------------------------------
\begin{equation}
	\left\langle \overline{ | M_{tot} |^2 } \right\rangle = \frac{ \left\langle\overline {| {\cal T}_{tot}|^2 } \right\rangle }{ \overline{ |{\cal T}_{el}|^2Ê} } =  \left\langle \overline{ | M_q^2 | } \right\rangle+ \left\langle \overline{ | M_g^2 | } \right\rangle+ 2 \text{Re}  \left\langle \{Ê\overline{ M_g M_q^\dagger } \} \right\rangle,
	\label{eq:contributions}
\end{equation}
where
%-----------------------------------
% Equation
%-----------------------------------
\begin{equation}
	{\cal T}_{el} = \bar{u} (p) M_h (p) \Rightarrow \overline{ |{\cal T}_{el}|^2 } = \sqrt{2} p_+ | M_h (p) |^2.
\end{equation}

As a consistency check we are able to recover the vacuum contribution in the limit of $x \rightarrow 1$ from the quark amplitude,
%-----------------------------------
% Equation
%-----------------------------------
\begin{equation}
	\left\langle \overline{ |M_q^2| } \right\rangle = \frac{ 2 g^2 C_F }{ \mathbf{q_\perp}^2 } x (1-x) \left\{ \frac{ 1 + (1-x)^2 }{ xÊ} \right\}
\end{equation}
%-----------------------------------
% Equation
%-----------------------------------
\begin{equation}
	\Rightarrow \left. x \frac{dI }{ dx d^2 \mathbf{k_\perp} } \right|_{x \rightarrow 1} \simeq \frac{ C_F \alpha_s }{ 2 \pi^2 } \frac{ 1 }{ \mathbf{k_\perp}^2 } = \frac{ \alpha_s }{ 2 \pi^2 } \frac{ 1 }{ \mathbf{k_\perp}^2 } P_{g \leftarrow q} (x \rightarrow 1)
\end{equation}
%-----------------------------------
% Equation
%-----------------------------------
%\begin{equation}
%	q_+ \frac{ dI^{vac} }{ dq_+ d^2 \mathbf{q_\perp} } = \frac{Ê\overline{ | M_{q}^{x \rightarrow 1 } |^2 } }{ 2 (2 \pi)^3 } = \frac{ \alpha_s C_FÊ}{ 2 \pi^2 } \frac{Êq_+}{ p_+ \mathbf{q_\perp^2 } } \Rightarrow \frac{ dI^{vac} }{ dz d^2 \mathbf{q_\perp} } = \frac{ \alpha_s }{ 2 \pi^2 } \frac{1}{\mathbf{q}_\perp^2} P^{vac}_{q \leftarrow q} (z \rightarrow 0)
%\end{equation}
with ${\bf k}_\perp=-{\bf q}_\perp$ and the vacuum splitting function \cite{Altarelli:1977zs,Dokshitzer:1977sg}
%-----------------------------------
% Equation
%-----------------------------------
\begin{equation}
	P^{vac}_{g \leftarrow q} (z) = C_F \left[ \frac{ 1 + (1-x)^2 }{ x } \right] \overset{x \rightarrow 1}{ \longrightarrow} C_F.
\end{equation}

\par As for the other two terms in equation \eqref{eq:contributions} (the medium contribution), the Dirac and color algebra are still  to be simplified. They can  be simplified using the polarization sum (with $\eta = (0, 1, \mathbf{0_\perp})$)
%the result is similar to the case $x \rightarrow 0$. The main differences lie in the  dependence on $x$ and in the color representation of the propagators $W$ and $G$, whose traces are now to be taken in the fundamental:
%-----------------------------------
% Equation
%-----------------------------------
\begin{equation}
	\sum_{\lambda} \epsilon_\mu^* (k, \lambda) \epsilon_\nu (k, \lambda) = -g_{\mu \nu} + \frac{ k_\mu \eta_\nu + k_\nu \eta_\mu }{ k \cdot \eta }
\end{equation}
and the relation between the Dirac spinors
%-----------------------------------
% Equation
%-----------------------------------
\begin{equation}
	\sum_{s} u (q, s)_\alpha \bar{u} (q, s)_\beta = \slashed{q}_{\alpha \beta} + m_{\alpha \beta}\,.
\end{equation}
 Using these two relations, we will end up with the trace of $\gamma$-matrices that are easily computed. For the color algebra, one can reduce all the traces to the fundamental representation using \cite{eikonalEvol}
%-----------------------------------
% Equation
%-----------------------------------
\begin{equation}
	W_{ab} (\mathbf{x_\perp}) = 2 \text{Tr} \left[ T^a W^F (\mathbf{x_\perp}) T^b W^{F\dagger} (\mathbf{x_\perp}) \right]
\end{equation}
to simplify the expression.

\par Putting all the kinematics in terms of the initial energy $p_+$ and the fraction of momentum carried away by the gluon, $x$, the medium amplitude can be written as:
%-----------------------------------
% Equation
%-----------------------------------
\begin{equation}
\begin{split}
	\left\langle \overline{ | M_{med} | ^2 } \right\rangle & = \left\langle \overline{ | M_g |^2 } \right\rangle + \ 2 \text{Re} \left\langle \overline{ M_g M_q^\dagger } \right\rangle =\\
	& = g^2 C_F \frac{ 1+ (1-x)^2 }{ x } \frac{1}{p_+} \text{Re} \left\{ \frac{ 1}{ (1-x) x p_+ } \int dy_+ d\bar{y}_+ d\mathbf{x_\perp} d\mathbf{\bar{x}_\perp} d\mathbf{z_\perp} \text{e}^{- i \mathbf{q}_\perp \cdot ( \mathbf{x_\perp} - \mathbf{\bar{x}_\perp} ) } \right. \\
	&\times  \left. \frac{ 1 }{N } \frac{ \partial }{ \partial \mathbf{y_\perp} } \text{Tr} \left< G (y_+, \mathbf{y_\perp} = \mathbf{0}_\perp; \bar{y}_+, \mathbf{z_\perp} | q_+ ) W^\dagger ( y_+, \bar{y}_+; \mathbf{0}_\perp ) \right> _F \right. \\
	&\cdot \left. \frac{ 1 }{N }  \frac{ \partial }{ \partial \mathbf{\bar{y}_\perp} } \text{Tr} \left< G^\dagger (\bar{y}_+, \mathbf{\bar{y}_\perp} = \mathbf{0}_\perp; L_+, \mathbf{x_\perp} | q_+ ) G( \bar{y}_+, \mathbf{z_\perp}; L_+; \mathbf{x_\perp} | q_+) \right>_F  \right. \\
	& \left. + \ 2 \ \frac{ \mathbf{q_\perp } }{ {\bf q}_\perp^2 } \cdot \int dy_+ d \mathbf{x_\perp} \text{e}^{-i \mathbf{q_\perp} \cdot \mathbf{x_\perp} } \right.  \\
	&\times \left. \frac{ 1 }{N } \frac{ \partial }{ \partial \mathbf{y_\perp } } \text{Tr} \left< G (y_+, \mathbf{y_\perp} = \mathbf{0}_\perp; L_+, \mathbf{x_\perp} | q_+ ) W^\dagger (y_+. L_+; \mathbf{0}_\perp ) \right>_F \right\},
\end{split}
\end{equation}
where the internal momenta were substituted by
%-----------------------------------
% Equation
%-----------------------------------
\begin{equation}
	\mathbf{q}_{1\perp} G_0 (y_+, \mathbf{y_{\perp}} = \mathbf{0}+\perp; x_{1+}'; \mathbf{x}_{1\perp} | q_+) =i \frac{ \partial }{ \partial \mathbf{y_{\perp}}} G_0 (y_+, \mathbf{y_{\perp}} = \mathbf{0}_\perp; x_{1+}'; \mathbf{x}_{1\perp} | q_+).
	%& = - i \boldsymbol{\epsilon_\perp}^* \cdot \frac{ \partial }{ \partial \mathbf{y_{\perp}}} G_0 (y_+, \mathbf{y_{\perp}} = \mathbf{0}_\perp; x_{1+}'; \mathbf{x}_{1\perp} | q_+).
\end{equation}
%and $\langle\cdots  \rangle$ denotes the medium average of Wilson lines and Green's functions.

\par Making the medium averages whose results can be taken from \cite{Yacine}, the medium amplitude for one gluon emission in the limit  $x \rightarrow 1$ is,
%-----------------------------------
% Equation
%-----------------------------------
\begin{equation}
\begin{split}
	\left\langle \overline{ | M_{med} | ^2 }\right\rangle & = \frac{ g^2}{ p_+ } P_{g \leftarrow q} (x) \ \text{Re} \left\{ \frac{ 1 }{ (1-x) x p_+ } \int dy_+ d\bar{y}_+ d \mathbf{x_\perp} \text{e}^{- i \mathbf{q_\perp} \cdot \mathbf{ x_\perp } } \text{e}^{- \frac{C_F}{2} \int_{\bar{y}_+}^{L_+} d\xi n(\xi) \sigma( \mathbf{x_\perp} ) } \right. \\
	&\times  \left. \frac{ \partial }{ \partial \mathbf{x_\perp} }\cdot  \frac{ \partial }{ \partial \mathbf{y_\perp} } \mathcal{K} (y_+, \mathbf{y_\perp} = \mathbf{0}_\perp; \bar{y}_+ \mathbf{x_\perp}|q_+ )  \right. \\
	& \left. +\  2 \frac{ \mathbf{q_\perp} }{ {\bf q}_\perp^2 }\cdot  \int dy_+ d\mathbf{x_\perp} \text{e}^{- i \mathbf{q_\perp} \cdot \mathbf{x_\perp} } \frac{ \partial }{ \partial \mathbf{y_\perp} } \mathcal{K} (y_+, \mathbf{y_\perp} = \mathbf{0}_\perp ; L_+, \mathbf{x_\perp} |q_+) \right\}.
	\label{eq:spc_x1}
\end{split}
\end{equation}

\par Although this expression was derived in the limiting case when $x \rightarrow 1$, we end up with an expression that depends explicitly on the fraction of energy carried by the emitted gluon. The result that we found for the medium amplitude is the exact vacuum splitting function, that appears as an overall pre-factor, corrected by two terms that depends on the medium parameters. Also, we have an explicit dependency with the initial energy that did not appeared in the limit of $x \rightarrow 0$. Moreover, if in this expression we take this limit, $x \rightarrow 0$, exchange the momenta of the path integral, $q_+$ by $k_+$, exchange the transverse momenta ${\bf q_\perp}$ by ${\bf k_\perp}$, and take the color representation in the adjoint one, we recover the previous results derived in the limit $x \rightarrow 0$ (see ref. \cite{lectures}). This indicates that it is possible to generalize this expression to include both limiting cases of the single gluon emission spectrum inside a medium. We will see a possible generalization in the next section.

\section{Multiple soft scattering approximation}
\label{MSS}

\par Being the general analytical solution of the path-integral from equation \eqref{eq:spc_x1} unknown, an approximation scheme must be used\footnote{A comparison of the results for the soft gluon spectrum obtained using different analytical approximations  and exact numerical solutions for the path integral, can be found in \cite{CaronHuot:2010bp}.}. One of the possible choices is  the dipole approximation. This is based in the observation that the Yukawa potential $a_0$ from the dipole cross section (eq. \eqref{eq:DipoleX}) has a leading quadratic dependency for small transverse distances $|\mathbf{r_\perp}|$. This allows to write the dipole cross section to logarithmic accuracy as \cite{Zakharov,Baier:1996sk}
%-----------------------------------
% Equation
%-----------------------------------
\begin{equation}
	n(\xi) \sigma(\mathbf{r_\perp}) \simeq \frac{1}{2} \hat{q} (\xi) \mathbf{r}_\perp^2\,,
\end{equation}
where $\hat{q} (\xi)$ is the transport coefficient which characterizes the transverse momentum squared , $\mu^2$, transferred from the medium to the projectile per mean free path, $\lambda$. This is the main parameter of this approximation and encodes all the dynamical properties of the medium. Generally, this parameter is time-dependent since the medium is expanding but, for a static medium, it can be written as $\hat{q} = \mu^2 / \lambda $.
%
%%-----------------------------------
%% Equation
%%-----------------------------------
%\begin{equation}
%	\hat{q} = \frac{ \mu^2 }{ \lambda }
%\end{equation}

In this paper, we study only the case of a static medium, since the results can be easily generalized to an expanding medium\footnote{The case of an expanding medium is just a generalization of equations \eqref{eq:osc_harm_1} and \eqref{eq:osc_harm_2} (see refs. \cite{lectures, ScalingLaw}).}. In this case, the path-integral becomes that of a two-dimensional harmonic oscillator \cite{Zakharov},
%-----------------------------------
% Equation
%-----------------------------------
\begin{equation}
\begin{split}
	\mathcal{K} (y_+, \mathbf{y_\perp} = \mathbf{0_\perp}; \bar{y}_+, \mathbf{x_\perp}|q_+ ) & = \mathcal{K}_{osc} (y_+, \mathbf{y_\perp} = \mathbf{0}_\perp; \bar{y}_+, \mathbf{x_\perp} |q_+)  \\
	& = \frac{ A_1 }{ \pi i } \exp \left[ i A_1 B_1 ({\bf x}_\perp^2 + {\bf y}_\perp^2 ) - 2 i A_1 \mathbf{x_\perp} \cdot \mathbf{y_\perp} \right],
\end{split}
\label{eq:osc_harm_1}
\end{equation}
%-----------------------------------
% Equation
%-----------------------------------
\begin{equation}
	A_1 = \frac{ q_+ \Omega }{ 2 \sin \left[ \Omega (\bar{y}_+ - y_+ ) \right] } \ \ , \ \ B_1 = \cos \left[ \Omega (\bar{y}_+ - y_+ ) \right] \ \ , \ \ \Omega = \frac{ 1-i}{2} \sqrt{ \frac{ \hat{q}_F }{ q_+ } }\ ,
	\label{eq:osc_harm_2}
\end{equation}
with imaginary frequency $\Omega$.

\par Note the subscript $F$  in the medium transport parameter to distinguish it from the usual transport parameter that is obtained from  the medium averages in the adjoint representation,
%-----------------------------------
% Equation
%-----------------------------------
\begin{equation}
	\frac{1}{N^2-1} \text{Tr} \left\langle W^A (\mathbf{x_\perp}) W^{\dagger A} (\mathbf{y_\perp}) \right\rangle = \exp \left\{ -\frac{C_A}{2} \int dx_+ \sigma (\mathbf{x_\perp} - \mathbf{y_\perp}) n(x_+) \right\}
\end{equation}
compared to the fundamental that is our case,
%-----------------------------------
% Equation
%-----------------------------------
\begin{equation}
	\frac{1}{N} \text{Tr} \left\langle W (\mathbf{x_\perp}) W^{\dagger} (\mathbf{y_\perp}) \right\rangle = \exp \left\{ -\frac{C_F}{2} \int dx_+ \sigma (\mathbf{x_\perp} - \mathbf{y_\perp}) n(x_+) \right\}.
\end{equation}
The two medium parameters are related by
%-----------------------------------
% Equation
%-----------------------------------
\begin{equation}
	\hat{q}_F = \frac{C_F}{C_A} \hat{q}\,.
\end{equation}

\par Substituting expressions \eqref{eq:osc_harm_1} and \eqref{eq:osc_harm_2} in \eqref{eq:spc_x1}, one obtains the medium spectrum within the dipole approximation for the limiting case $x \rightarrow 1$:

\begin{equation}
	q_+ \frac{ dIÊ}{ dq_+ d \mathbf{q_\perpÊ}^2} = \frac{Ê\left\langle \overline{M_{med}} \right\rangle }{ 4 (2\pi)^2 }\  = \frac{ \alpha_sÊ}{ 4\pi } P_{g \leftarrow q} (x) (I_1 + I_2 ),
	\label{eq:DDspec1}
\end{equation}
with
%-----------------------------------
% Equation
%-----------------------------------
\begin{equation}
\begin{split}
	I_1 & = \frac{ 1 }{ x (1-x) p_+^2 } \text{Re}  \int dy_+ d\bar{y}_+ d \mathbf{x_\perp} \text{e}^{- i \mathbf{q_\perp} \cdot \mathbf{ x_\perp } } \text{e}^{- \frac{{C_F}}{2} \int_{\bar{y}_+}^{L_+} d\xi n(\xi) \sigma( \mathbf{x_\perp} ) } \\
	&\times \frac{ \partial }{ \partial \mathbf{x_\perp} }\cdot \frac{ \partial }{ \partial \mathbf{y_\perp} } \mathcal{K} (y_+, \mathbf{y_\perp} = \mathbf{0}_\perp; \bar{y}_+ \mathbf{x_\perp}|q_+ ) \\ 
	& = \frac{ 1 }{ x (1-x) p_+^2 } \times \text{Re} \left\{ \int_{0}^{L_+} dy_+ \int_{y_+}^{L_+} d\bar{y}_+  \right. \\
	& \left. \left[ \frac{ -2 A_1^2 D }{ (D-i A_1 B_1 )^2 } + \frac{ i A_1^3 B_1 }{ 2 (D-i A_1 B_1)^3 } {\bf q}_\perp^2 \right] \exp \left[ \frac{ - {\bf q}_\perp^2 }{ 4 (D-iA_1 B_1 ) } \right] \right\}
	\end{split}
\end{equation}
representing the \textit{gluon} term ($\left\langle \overline{ | M_g|^2 } \right\rangle$) and
%-----------------------------------
% Equation
%-----------------------------------
\begin{equation}
\begin{split}
	I_2 & = \frac{2}{ p_+} \frac{ \mathbf{q_\perp} }{ {\bf q}_\perp^2 }\cdot  \int dy_+ d\mathbf{x_\perp} \text{e}^{- i \mathbf{q_\perp} \cdot \mathbf{x_\perp} } \frac{ \partial }{ \partial \mathbf{y_\perp} } \mathcal{K} (y_+, \mathbf{y_\perp} = \mathbf{0}_\perp ; L_+, \mathbf{x_\perp} |E_+) \\
	& = \frac{ 2 }{ p_+ } \text{Re} \left\{ \int_0^{L_+} dy^+ \frac{-i}{B_2^2 } \exp \left[ -i \frac{ {\bf q}_\perp^2 }{ 4 A_2 B_2 } \right] \right\}
\end{split}
\end{equation}
the \textit{interference} term. The variables $A_1$, $B_1$ and $\Omega$ are the same as in equation \eqref{eq:osc_harm_2}, while $A_2$, $B_2$ and $D$ read
\begin{equation}
%\begin{gathered}
%	A_1 = \frac{ q_+ \Omega }{ 2 \sin \left[ \Omega (\bar{y}_+ - y_+ ) \right] } \ \ , \ \ B_1 = \cos \left[ \Omega (\bar{y}_+ - y_+ ) \right] \ \ , \ \ \Omega = \frac{ 1-i}{2} \sqrt{ \frac{ \hat{q}_F }{ q_+ } }\ , \\
	A_2 = \frac{ q_+ \Omega }{ 2 \sin \left[ \Omega (L_+ - y_+ ) \right] } \ \ , \ \ B_1 = \cos \left[ \Omega (L_+ - y_+ ) \right] \ \ , \ \ \ D = \frac{1}{4} (L_+ - \bar{y}_+) \hat{q}_F.
	\label{eq:ABDx1}
%\end{gathered}
\end{equation}

\subsection{Interpolation between the hard ($x \rightarrow 1$) and soft ($x \rightarrow 0$) gluon emission }

\par To provide a heuristic interpolation between both limits, we should decide which variables we want for analyzing the spectrum since we have available $q_+$, $p_+$ and $x$, but they are constrained by the relation $q_+ = (1-x) p_+$. The most direct single inclusive spectrum that we could compute from expression \eqref{eq:spc_x1}, should be the one related to the final quark, as previously derived. But since we want to find an interpolation function that also describes the single inclusive spectrum for the gluon, a good observable is the energy loss, independent of which parton in the splitting carries a larger fraction of the initial energy. Thus, if we substitute
\begin{equation}
	q_+ = (1-x) p_+ \rightarrow x (1-x ) p_+ \equiv E_+,
	\label{eq:sub1}
\end{equation}
when $x \rightarrow 0$, $E_+ \rightarrow x p_+ = k_+$ and, when $x \rightarrow 1$, $E_+ \rightarrow (1-x) p_+ = q_+$. This means that
\begin{equation}
\begin{split}
	q_+ \frac{ dI }{ dq_+ d \mathbf{q_\perp}^2 } \rightarrow x (1-x) p_+ \frac{ dI }{ d [ x (1-x) p_+] d \mathbf {q_\perp}^2 } & \rightarrow k_+ \frac{ dI }{ d k_+  d \mathbf{q_\perp}^2 } \text{ for } x \rightarrow 0. \\
	& \rightarrow q_+ \frac{ dI }{ dq_+ d \mathbf{q_\perp}^2 } \text{ for } x \rightarrow 1.
\end{split}
\label{interpol}
\end{equation}

\par The transverse momenta
%has a minus sign between the two cases, but since it appears squared, it is not a problem. It 
are constrained by
\begin{equation}
	\mathbf{q_\perp}^2 \leq 2 x^2 (1-x)^2 p_+^2\,.
\end{equation}
As for the medium parameter, $\hat{q}_F$, we can substitute it by \cite{Zakharov:1997uu,Arnold:2009mr}
\begin{equation}
	\hat{q} = x \hat{q}_F + (1-x) \hat{q}_A = \frac{ x C_F + (1-x) C_A }{ C_A } \hat{q}_A.
	\label{eq:sub2}
\end{equation}

\par With all these considerations, we are able to describe both limits within the considered approximations and recover the results that were derived in this manuscript and in  \cite{lectures}. So, applying the above formulas to equation \eqref{eq:DDspec1} we get
%-----------------------------------
% Equation
%-----------------------------------
\begin{equation}
	E_+ \frac{ dI^{med} }{ dE_+ d{\bf q}_\perp^2 } = \frac{Ê\alpha_s C_F }{ 4 \pi } x (1-x) P_{g \leftarrow q} (x) (I_1 + I_2)
	\label{eq:spec_ocs_harm}
\end{equation}
with
%-----------------------------------
% Equation
%-----------------------------------
\begin{equation}
\begin{split}
	I_1 & = \frac{ 1 }{ E_+^2 } \text{Re}  \int dy_+ d\bar{y}_+ d \mathbf{x_\perp} \text{e}^{- i \mathbf{q_\perp} \cdot \mathbf{ x_\perp } } \text{e}^{- \frac{1}{2} \int_{\bar{y}_+}^{L_+} d\xi n(\xi) \sigma( \mathbf{x_\perp} ) } \\
	&\times \frac{ \partial }{ \partial \mathbf{x_\perp} }\cdot \frac{ \partial }{ \partial \mathbf{y_\perp} } \mathcal{K} (y_+, \mathbf{y_\perp} = \mathbf{0}_\perp; \bar{y}_+ \mathbf{x_\perp}|E_+ ) \\ 
	& = \frac{ 1 }{ E_+^2 } \times \text{Re} \left\{ \int_{0}^{L_+} dy_+ \int_{y_+}^{L_+} d\bar{y}_+  \right. \\
	& \left. \left[ \frac{ -2 A_1^2 D }{ (D-i A_1 B_1 )^2 } + \frac{ i A_1^3 B_1 }{ 2 (D-i A_1 B_1)^3 } {\bf q}_\perp^2 \right] \exp \left[ \frac{ - {\bf q}_\perp^2 }{ 4 (D-iA_1 B_1 ) } \right] \right\},
	\end{split}
\end{equation}
and
%-----------------------------------
% Equation
%-----------------------------------
\begin{equation}
\begin{split}
	I_2 & = \frac{2}{ E_+} \frac{ \mathbf{q_\perp} }{ {\bf q}_\perp^2 }\cdot  \int dy_+ d\mathbf{x_\perp} \text{e}^{- i \mathbf{q_\perp} \cdot \mathbf{x_\perp} } \frac{ \partial }{ \partial \mathbf{y_\perp} } \mathcal{K} (y_+, \mathbf{y_\perp} = \mathbf{0}_\perp ; L_+, \mathbf{x_\perp} |E_+) \\
	& = \frac{ 2 }{ E_+ } \text{Re} \left\{ \int_0^{L_+} dy^+ \frac{-i}{B_2^2 } \exp \left[ -i \frac{ {\bf q}_\perp^2 }{ 4 A_2 B_2 } \right] \right\}.
\end{split}
\end{equation}
The variables $A_1$, $A_2$, $B_1$, $B_2$, $\Omega$ and $D$ are the same as in equation \eqref{eq:ABDx1} but with the substitutions \eqref{eq:sub1} and \eqref{eq:sub2}:
\begin{equation}
\begin{gathered}
	A_1 = \frac{ E_+ \Omega }{ 2 \sin \left[ \Omega (\bar{y}_+ - y_+ ) \right] } \ \ , \ \ B_1 = \cos \left[ \Omega (\bar{y}_+ - y_+ ) \right] \ \ , \ \ \Omega = \frac{ 1-i}{2} \sqrt{ \frac{ \hat{q} }{ E_+ } }\ , \\
	A_2 = \frac{ E_+ \Omega }{ 2 \sin \left[ \Omega (L_+ - y_+ ) \right] } \ \ , \ \ B_1 = \cos \left[ \Omega (L_+ - y_+ ) \right] \ \ , \ \ \ D = \frac{1}{4} (L_+ - \bar{y}_+) \hat{q}.
	\label{eq:ABDInterp}
\end{gathered}
\end{equation}

Note that while for either $x\to 0$ or $x\to 1$, the meaning of ${\bf q}_\perp$ in Eq. (\ref{interpol}) is clear as the transverse momentum of the emitted soft gluon (where ${\bf q}_\perp=-{\bf k}_\perp$) or quark respectively, the interpretation for intermediate $x$ is far more involved. Actually, such kinematical situation requires the computation of a double inclusive cross section in which the Brownian motion of both outgoing partons is considered. We leave this computation for future studies.

\section{Numerical Results}
\label{sec:results}

\par
\begin{figure}[htbp]
	\begin{center}
		\subfigure[]{
			\includegraphics[width=1.0\textwidth]{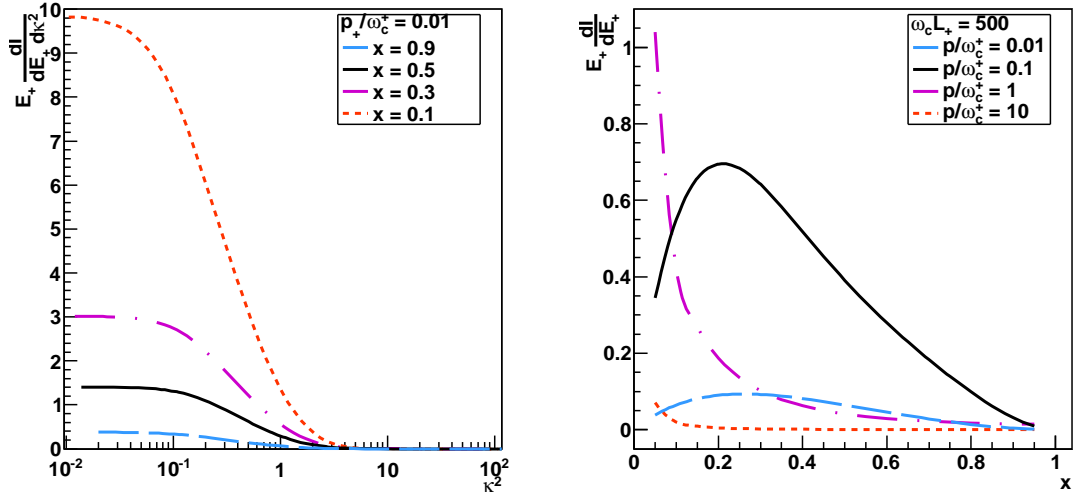}
			\label{fig:spec1}
		}
		\subfigure[]{
			\includegraphics[width=1.0\textwidth]{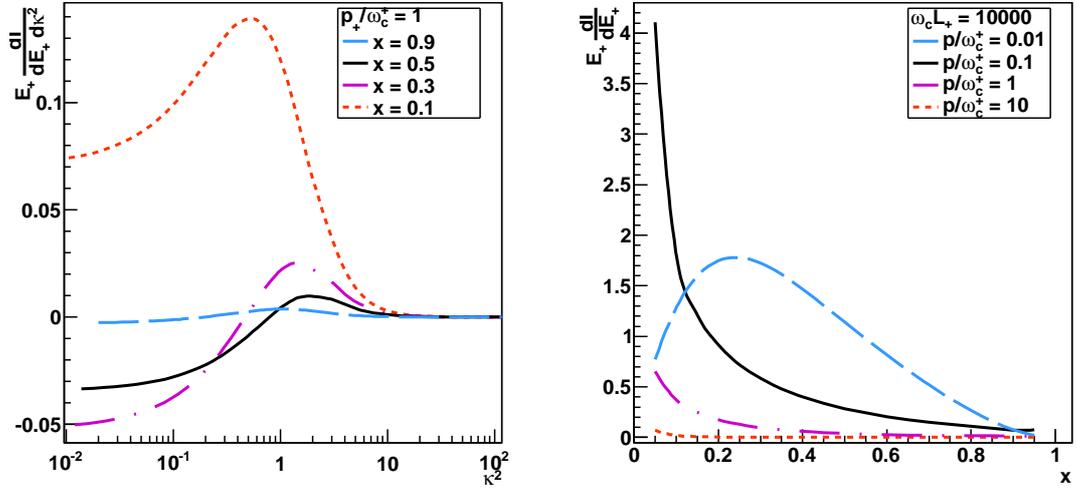}
			\label{fig:spec2}
		}
		\caption{Evolution of the medium-induced gluon radiation spectrum \eqref{eq:spec_ocs_harm} of a quark in a static medium for different values of $x$ and $p_+/\omega_c^+$. The left plots correspond to the double differential spectrum (two different values of $p_+/ \omega_c^+$ in (a) and (b)) and the right ones are the same integrated over $\mathbf{q_\perp}$ (two different values of $\omega_c^+ L_+$ in (a) and (b)).}
		\label{fig:spec}
	\end{center}
\end{figure}

 The results for the double-differential medium-induced gluon radiation spectrum for a quark traversing a static medium in the limit of hard gluon emission are presented in figure \ref{fig:spec}. The results are plotted as  functions of the following dimensionless variables (see \cite{ASW1,ASW2} for the corresponding variables in the case of soft gluon emissions):
%-----------------------------------
% Equation
%-----------------------------------
\begin{equation}
	\omega_c^+ = \frac{1}{2} \hat{q_A} L_+^2 \ \ \ , \ \ \ \ \frac{ p_+ }{ \omega_c^+ } = \frac{ 2 p_+ }{ \hat{q}_A L_+^2 } \ \ \ \ , \ \ \ \ \kappa^2 = \frac{ q_\perp^2 }{ \hat{q}_A L_+ } \ \ \ , \ \ \ x \ .
	\label{eq:var}
\end{equation}
The left plots correspond to the double differential spectrum. By integrating out $\mathbf{q_\perp}$, we  obtain the right-handed ones (the analytical expressions for this integration in the limit $\omega_c L_+ \rightarrow \infty$ are written in appendix \ref{appB}). Figure \ref{fig:spec1} presents the results for small medium parameters and figure \ref{fig:spec2} for larger  ones.
%***********************************
% Image 
%***********************************

\par The two asymptotic behaviors that one  expects are confirmed: when the energy or the transverse momentum of the emitted parton increase, the spectrum goes to zero. Another feature of the spectrum is the suppression in the low $\kappa^2$ region. For values $p_+ / \omega_c \simeq 1$ for every value of $x$, the spectrum becomes negative - the total radiation is smaller than in vacuum -, which does not happens when $p_+ / \omega_c^+$ is smaller. For the case of the soft gluon emission, this suppression can be understood in terms of the formation time of the outgoing gluon:
%-----------------------------------
% Equation
%-----------------------------------
\begin{equation}
	t_{form}^g \simeq \frac{ 2k_+ }{ q_\perp^2}\,.
\end{equation}

\par In the present case, we can generalize this concept to include both final particles and define the formation time of the outgoing parton as
%-----------------------------------
% Equation
%-----------------------------------
\begin{equation}
	t_{form} \simeq \frac{ 2E_+ }{ q_\perp^2}\,.
\end{equation}
 In the regime where $t_{form} \gg L_+$, the function becomes highly oscillatory and we enter in the regime of the Landau-Pomeranchuk-Migdal (LPM) effect \cite{LPM1, Zakharov}. As a result, a suppression of the spectrum due to the inefficiency of exchanging information with the medium is observed. In the upper left plot this behavior is less striking since we increase the medium length by two orders of magnitude.

\section{Conclusions and Outlook}
\label{sec:conclusions}

\par Summarizing, we have computed the medium-induced gluon radiation spectrum in the limit of a hard emission. Making use of the path-integral formalism to describe the propagation of high-energy particles inside a medium in terms of Wilson lines and Green's functions, we found a finite-energy correction to the double differential spectrum of the final quark. Since we are able to recover the soft gluon emission limit with the expression derived in this manuscript, we provide a generalization that is able to describe both limiting cases. The result for this interpolation function (shown in equation \eqref{eq:spec_ocs_harm} for the multiple soft scattering approximation) contains, as in the case of soft gluon emissions, two contributions for the medium spectrum: the gluon term, which is suppressed by the initial momenta, and the interference term. The spectrum vanishes, as expected, in the kinematical limit.
%previous results of the \cite{ASW}

\par Applying the multiple soft scattering approximation, we obtained explicit expressions for the case of a static medium. The integrals in the longitudinal variables were performed numerically and the main results were shown in figure \ref{fig:spec}. The conclusions are a clear suppression of the spectrum for small values of the parton transverse momentum, as predicted by the LPM effect. Also, the density of the medium constrains strongly the possible energy range of radiation that is emitted.

\par Finite-energy corrections are a key ingredient for the quantitative description of the energy loss processes in whatever used formalism. They are automatically included in Monte Carlo approaches, though until now in a heuristic form, see e.g.  \cite{qpythia1, qpythia2,Zapp:2012nw} for a discussion of its implementation and impact in a Monte Carlo framework. From this work we found a generalization that is still half-way of its true form: although we have an expression that is valid for all fraction of energies of the emitted parton, there is still one parton (the hardest one) that has its movement constrained in the transverse plane (this one is always described by a Wilson line). The outlook of this work is the implementation of these corrections in a Monte Carlo code in order to study its phenomenological consequences since now we are able to access the intermediate region of $x$, and, on the other hand, the full generalization of this result through allowing both final particles to have a Brownian motion in the transverse plane.

\section*{Acknowledgments}

\par We thank Y. Mehtar-Tani and J. G. Milhano for many discussions during the elaboration of this paper. This work is  supported by the European Research Council grant HotLHC ERC-2001-StG-279579; by Ministerio de Ciencia e Innovaci\'on of Spain grants FPA2008-01177, FPA2009-06867-E and Consolider-Ingenio 2010 CPAN CSD2007-00042; by Xunta de Galicia grant PGIDIT10PXIB \-206017PR (LA and NA);  by Funda\c c\~ao para a Ci\^encia e a Tecnologia
of Portugal under projects SFRH/BD/64543/2009 and CERN/FP/116379/2010 (LA); and by FEDER. CAS is a Ram\'on y Cajal researcher.

\appendix

\section{Gamma matrices}
\label{appA}

In the light-cone the gamma matrices read
\begin{subequations}
	\begin{gather}
		\gamma^+ = \frac{\gamma^0 + \gamma^3}{\sqrt{2}} = \frac{1}{\sqrt{2}} \left(
		\begin{array}{cc}
			\mathbf{1} & \sigma^3 \\
			- \sigma^3 & \mathbf{- 1}
		\end{array}
		\right),
		\\
		\gamma^- = \frac{\gamma^0 - \gamma^3}{\sqrt{2}} = \frac{1}{\sqrt{2}} \left(
		\begin{array}{cc}
			\mathbf{1} & - \sigma^3 \\
			\sigma^3 & \mathbf{- 1}
		\end{array}
		\right),
		\\
		\gamma^1 = \left(
		\begin{array}{cc}
			0 & \sigma^1 \\
			-\sigma^1 & 0
		\end{array}
		\right),
		\\
		\gamma^2 = \left(
		\begin{array}{cc}
			0 & \sigma^2 \\
			-\sigma^2 & 0
		\end{array}
		\right).
	\end{gather}
\end{subequations}

They obey to the following relations:
\begin{equation}
	\begin{array}{cccc}
		\begin{array}{c}
			(\gamma^+)^\dagger = \gamma^- \,,\\
			(\gamma^-)^\dagger = \gamma^+ \,,\\
			(\boldsymbol{\gamma_\perp})^\dagger = - \boldsymbol{\gamma_\perp}\,,
		\end{array} &
		\begin{array}{c}
			\boldsymbol{\gamma_\perp} \cdot \boldsymbol{\gamma_\perp} = -2 \,,\\
			\gamma^+ \gamma^+ = 0 \,,\\
			\gamma^- \gamma^- = 0\,,
		\end{array} & 
		\begin{array}{c}
			\gamma^+ \gamma^- \gamma^+ = 2 \gamma^+\,, \\
			\gamma^- \gamma^+ \gamma^- = 2 \gamma^- \,,\\
			\boldsymbol{\gamma_\perp} \gamma^- \boldsymbol{\gamma_\perp} = 2 \gamma^-\,,
		\end{array} &
		\begin{array}{c}
			\boldsymbol{\gamma_\perp} \gamma^+ \boldsymbol{\gamma_\perp} = 2 \gamma^+ \,,\\
			\gamma^+ \boldsymbol{\gamma_\perp} \gamma^+ = 0 \,,\\
			\gamma^- \boldsymbol{\gamma_\perp} \gamma^- = 0\,.
		\end{array}
	\end{array}
\end{equation}

\section{BDMPS limit}
\label{appB}

\par From expression \eqref{eq:spec_ocs_harm}, we are able to compute the BDMPS limit, taking the limit $R = \omega_c L_+ \rightarrow \infty$. This is equivalent to performing the integration in the transverse momentum taking into account an opening angle for the parton emission, $\Theta$, $0 < q_\perp < \chi E_+$, where $\chi = \sin \Theta$ and, then, take the limit $\chi \rightarrow \infty$. Doing the integration one finds that
\begin{equation}
	E_+ \frac{ dI^{med} }{ dE_+ } = \frac{ \alpha_s }{ 4 \pi } (1-x) x P_{g \leftarrow q} (x) (I_1+ I_2),
\end{equation}
where
\begin{equation}
	I_1 = \frac{ 8 A_1^2 }{ E_+ ^2 } \left\{ -1 + \exp \left[ - \frac{ (\chi E_+ )^2 }{ 4 (D - i A_1 B_1 ) } \right] \left( 1 - \frac{  i A_1 B_1 (\chi E_+ )^2 }{ 4 (D - i A_1 B_1 )^2 } \right) \right\},
\end{equation}
\begin{equation}
	I_2 = \frac{ 8 A_2 }{ B_2 E_+ } \left\{  \exp \left[ - \frac{ i (\chi E_+ )^2 }{ 4 A_2 B_2 } \right] -1\right\}
\end{equation}
and $A_1$, $A_2$, $B_1$, $B_2$, $\Omega$ and $D$ are defined in equation \eqref{eq:ABDInterp}.

\par Taking the limit $\chi \rightarrow \infty$, the previous equations can be reduced to
\begin{equation}
	\lim_{R \rightarrow \infty} I_1 = 2 \text{Re} \int_{0}^{L_+} dy_+ \frac{ \Omega \cos (\Omega y_+ ) }{ \sin ( \Omega y_+ ) }\,,
\end{equation}
\begin{equation}
	\lim_{R \rightarrow \infty } I_2 = - 2 \text{Re} \int_{0}^{L_+} dy_+ \frac{ \Omega }{ \cos (\Omega y_+ ) \sin (\Omega y_+) }\,.
\end{equation}
 Both integrals are divergent but, when summing the two contributions, the integral over the remaining longitudinal coordinate is finite and the the result is given by:
\begin{equation}
	E_+ \frac{ dI^{med} }{d E_+} = \frac{ \alpha_s }{ 2 \pi} x (1-x) P_{g \leftarrow q} (x) \text{Re} \ln \left[ \cos (\Omega L_+ ) \right].
\end{equation}
This result agrees with the ones previously derived when doing the limit $x \rightarrow 1$ and $x \rightarrow 0$ respectively.

%\bibliographystyle{ieeetr}
%\bibliography{Bibliography2}

\end{document}